\documentstyle[aps,prb,epsf,multicol]{revtex}
\begin{document}
\title{Practical Methods for Ab Initio Calculations on Thousands of Atoms}
\author{D.R.Bowler\protect\cite{UCL}}
\address{Department of Physics and Astronomy, University College London, 
Gower Street, London, WC1E 6BT, UK}
\author{I.J.Bush}
\address{DCI, CLRC Daresbury Laboratory, Warrington, Cheshire}
\author{M.J.Gillan\protect\cite{DCI}}
\address{Department of Physics and Astronomy, University College London,
Gower Street, London, WC1E 6BT, UK}
\maketitle
\begin{abstract}
We describe recent progress in developing practical {\it ab initio} methods
for which the computer effort is proportional to the number of atoms: linear 
scaling or ${\cal O} (N)$ methods.  It is shown that the locality property of
the density matrix gives a general framework for constructing such methods.
We then describe our scheme, which operates within density
functional theory.  Efficient methods for reaching the electronic 
ground state are summarised, both for finding the density matrix, 
and in varying the localised
orbitals.  
\end{abstract}

\begin{center}
{\bf Submitted to the International Journal of Quantum Chemistry (1999)}
\end{center}

\begin{multicols}2
\section{Introduction}
Over the last thirty years, the use of {\it ab initio} electronic structure
techniques has become widespread in chemistry and physics.
However, all traditional techniques are limited in the system size 
which they can treat (often defined by the number of atoms, $N$) 
by poor scaling of the computational effort required, which is 
generally at least ${\cal O}(N^2)$, if not
worse.  In the context of standard self-consistent field (SCF) quantum 
chemical methods such as Hartree-Fock theory or density functional theory
(DFT), there are two demanding parts of
a calculation: first, the build of the Fock matrix (or, in DFT, the
Hamiltonian matrix), which can scale as ${\cal
O}(N^2)$; second, the solution for eigenvectors of the Fock matrix, which
scales as ${\cal O}(N^3)$ if performed as a diagonalisation; this only
dominates for large values of $N$.  
An alternative to diagonalisation is iterative minimisation,
which has ${\cal O}(N^2)$ scaling (one $N$
dependence comes from the eigenvectors spreading over all space, and the other
from the number of eigenvectors, which depends on $N$); if, as is often the
case, the
eigenvectors must be orthogonalised to each other, this leads asymptotically
to ${\cal O}(N^3)$ scaling.  Whatever the cause of poor scaling,
it results in a practical limit of a few hundred atoms in
conventional {\it ab initio} techniques.  The desire to model large systems,
e.g. biomolecules or nanostructures, has seen a strong push in recent years to
achieve linear scaling with system size.

The building of the Fock matrix (and specifically the Coulomb and exchange
terms which are the most expensive) with linear scaling has been addressed
recently by several groups\cite{Ochsen98,Schweg96,Burant96}; however, this 
aspect will not concern us here.  Rather, we will consider linear scaling 
techniques (which also rely on iterative minimisation) for finding the 
self-consistent ground state of the system.

Recently, many linear scaling techniques have been proposed, which are all 
based on the search for the density 
matrix\cite{(Pettifor 1989),Yang91,(Galli 1992),%
(Li etal 1993),(Daw 1993),(Ordejon etal 1993),(Aoki 1993),mauri93,%
(Goedecker 1994),(Stechel et al 1994),hierse94,kohn95,Her95,%
(Kress and Voter 1995),(Horsfield 1996),(Horsfield etal 1996),kohn96,%
Hern96,Goringe97,baer97,haynes97}.  They start from the observation
that the density matrix between two points in space decays in some manner as
those two points increase in separation.  This is intuitively clear, as it is
well-known that bonding is local.  The result is that the electronic structure
of an atom depends only on its {\it local} environment, so that if the overall
size of the system changes, there is no effect on the local electronic
structure; thus the effort required to solve for the whole system should be
proportional to $N$.  This is the foundation of all
linear scaling electronic structure methods (with a few
exceptions\cite{Madden94,Wang94} which will not concern us here).

The paper is divided up as follows: Section~\ref{ONTh} describes 
the basic theory behind our ${\cal O}(N)$ DFT method, while 
Section~\ref{Strat} presents recent advances we have made in searching for the
electronic ground state.  The future directions the work will take are shown
in Section~\ref{Fut}, and the paper is concluded in Section~\ref{Conc}.

\section{${\cal O}(N)$ density functional theory}
\label{ONTh}

The density matrix within density functional theory (DFT) can be written as:
\begin{equation}
\rho({\bf r},{\bf r^\prime}) = \sum_i f_i \psi_i({\bf r}) 
\psi_i^*({\bf r}^\prime),
\end{equation}
where $\psi_i$ is a Kohn-Sham eigenfunction, and $f_i$ is the occupancy of
that eigenfunction.  
The key observation which underpins ${\cal O}(N)$ DFT is that DFT can be
formulated in terms of $\rho({\bf r},{\bf r^\prime})$, and that the ground 
state can be found by minimising the total energy, $E_{\rm tot}$, with respect
to $\rho({\bf r},{\bf r^\prime})$ subject to the condition that $\rho({\bf
r},{\bf r^\prime})$ is idempotent (these statements are proved 
elsewhere\cite{Hern96}).  Idempotency means that $\rho \cdot \rho =
\rho$, which is equivalent to the eigenvalues of $\rho$ (which are the
occupation numbers $f_i$) being either zero or
one.  This is a crucial property for the density matrix, as it
is ensures that the density matrix is a projector -- it is the operator
which projects onto the space of occupied states.

Another important property of the density matrix is that it decays as 
the separation between points increases:
\begin{equation}
\rho({\bf r},{\bf r}^\prime) \rightarrow 0 \ {\rm as} \mid {\bf r} - {\bf
r}^\prime \mid \rightarrow \infty.
\end{equation}
The fundamental reason for this decay is the loss of quantum phase coherence
between distant points.  The net result is that the local environment is all
that is important in determining $\rho({\bf r},{\bf r^\prime})$, and thus that
the amount of information contained in $\rho$ is proportional to $N$.  This 
decay property of the density matrix can be used to make the amount of 
information in the system strictly linear with the system size by 
{\it imposing} a constraint, and setting the density matrix to zero beyond a 
cutoff:
\begin{equation}
\rho({\bf r},{\bf r}^\prime) = 0, \mid {\bf r} - {\bf r}^\prime \mid > R_c,
\end{equation}
where $R_c$ is some cut-off radius.  Solving for the energy with this
constraint imposed, and that of idempotency,
will lead to an upper bound on the ground state energy\footnote{The
idempotency constraint on $\rho$ will make it the projection operator onto the
occupied states.  The energy of the ground state given by these states will 
be the energy for the system {\it under the constraint of a localised $\rho$}.  
Since DFT is variational, this extra constraint will raise the energy above
the true ground state energy (i.e. without localisation), and so will give
an upper bound to the true ground state energy.}; as
$R_c$ is increased, it will converge to the true ground state.  Clearly there
is a balance to be struck between accuracy, which increases as $R_c$ is
increased,  and the complexity of the computational problem (e.g. number of
variational degrees of freedom associated with $\rho$), which also increases
with $R_c$.

This exact formulation cannot be followed for practical methods, as $\rho$ is
dependent on two vector positions.  Instead, a further approximation is
introduced, namely that $\rho$ be {\it separable}, so that it can be written
in the form:
\begin{equation}
\rho({\bf r},{\bf r}^\prime) = \sum_{i\alpha,j\beta} 
\phi_{i\alpha}({\bf r}) K_{i\alpha,j\beta} \phi_{j\beta}({\bf r}^\prime),
\end{equation}
which is equivalent to requiring that $\rho$ only have a finite number of
non-zero eigenvalues.  The functions 
$\phi_{i\alpha}({\bf r})$ are known as `localised
orbitals', where $i$ runs over atoms and $\alpha$ over localised orbitals on
each atom.  The matrix $K_{i\alpha,j\beta}$ is the density matrix in the basis
set of $\{\phi_{i\alpha}({\bf r})\}$ (and is identical to the density
matrix commonly seen in ${\cal O}(N)$ tight binding schemes; indeed many of
these schemes can be used to solve for this matrix within ${\cal O}(N)$ DFT).

The localisation of $\rho$ is accomplished by setting the localised orbitals
to be non-zero only inside a certain radius, $R_{\rm reg}$, centred on the
atoms $i$.  A spatial cutoff (not generally of the same value) must also be
imposed on the matrix $K$, so that $K_{i\alpha,j\beta}=0$ 
once atoms $i$ and $j$ are more than a certain distance apart.

\subsection{A Specific Implementation: CONQUEST}
\label{sec:CQ}

The framework described above is completely general; we will now concentrate
on our specific implementation,
\textsc{Conquest}\cite{Her95,Hern96,Goringe97,Gillan98}.
This is based on the pseudopotential approach to DFT (described briefly in 
Appendix~\ref{sec:Pseudo}),
and has been constructed so as to be as accurate as conventional {\it ab
initio} pseudopotential calculations, which use plane waves as a basis set.

In practice, there are various issues which must be addressed: minimising the
total energy with respect to $K_{i\alpha,j\beta}$ while maintaining
idempotency and spatial localisation; representing the localised orbitals;
the cutoffs required to achieve good convergence to the true ground state; and
practical questions, including integration and implementation on parallel
computers.  These have been addressed in detail
elsewhere\cite{Goringe97,Gillan98}; we present a brief summary here.

The imposition of idempotency on $K$ during the minimisation of 
$E_{\rm tot}$ with respect to $K_{i\alpha,j\beta}$
is the hardest constraint to maintain.  There are several proposed means of
accomplishing this; the method described here is based on the purification
technique of McWeeny\cite{McWeeny60}, recently used 
in tight binding calculations by Li, Nunes and 
Vanderbilt\cite{(Li etal 1993)}, and described in detail in
Section~\ref{sec:DMM}.  
It requires $K$ to be written in terms of an `auxiliary'
density matrix, $L$:
\begin{eqnarray}
K = 3LSL - 2LSLSL,
\end{eqnarray}
where $S$ is the overlap matrix:
\begin{equation}
S_{i\alpha,j\beta} = \int {\rm d}{\bf r} \phi_{i\alpha}({\bf r})
\phi_{j\beta}({\bf r}).
\end{equation}
The localisation of $K$ is then imposed as a 
spatial cutoff on $L$:
\begin{equation}
L_{i\alpha,j\beta} = 0, \ \ \ \mid {\bf R}_{i} - {\bf R}_{j} \mid 
> R_{\rm L},
\end{equation}
where ${\bf R}_{i}$ is the position of atom $i$ and $R_{\rm L}$ is a 
cutoff radius.  The energy is then minimised with respect to the matrix
elements $L_{i\alpha,j\beta}$ using the standard conjugate gradients
technique\cite{NR}.  The effect of the purification is to make $K$ more
nearly idempotent given an $L$ which is nearly idempotent.

The next issue to consider is the representation of the localised orbitals.
The lessons learnt from the use of plane waves in 
conventional pseudopotential calculations are helpful to remember here.
First, they offer a systematic convergence of the energy 
with respect to the basis set completeness, and this is achieved with a single
parameter (the cutoff energy).  Second, they are bias free - that is, they are
completely flexible, and no knowledge of the kind of bonding in the system is
required.  If possible, the choice of basis set would reflect these qualities.

\textsc{Conquest} represents the localised orbitals in a real-space basis.
There are various possibilities for an efficient, real-space basis.  One 
is to use the spherical equivalent of plane waves, that is
spherical Bessel functions $j_l({\bf r})$ combined with spherical harmonics
$Y^l_m({\bf r})$, within each localisation region.  This representation has been
discussed by Haynes and Payne\cite{Haynes97}, but practical results have not
yet been reported.  Another ${\cal O}(N)$ DFT scheme\cite{Ord96} uses
pseudo-atomic orbitals\cite{SN} as the basis functions, with considerable
success.  An alternative is to represent the $\phi_{i\alpha}$ by
their values on a grid, and to calculate matrix elements of the kinetic energy
by taking finite differences.  This technique is well established in
conventional first principles calculations\cite{Chel94}, and has been
investigated in the context of ${\cal O}(N)$ techniques by us\cite{Her95} and
recently by Hoshi and Fujiwara\cite{Hoshi95}.  At present we use a basis of
B-splines, $\Theta({\bf r})$, (also called blip functions) in the expansion:
\begin{equation}
\phi_{i\alpha}({\bf r}) = \sum_s b_{i\alpha s} \Theta({\bf r}-{\bf R}_{is}),
\end{equation}
where the B-splines are piecewise polynomial functions (continuous up to the
third derivative) which are strictly localised on the points of a grid
(notated as ${\bf R}_{is}$ above) which is rigidly attached to each atom 
(known as the blip grid).  The energy is then minimised with respect to the 
coefficients of the B-splines, $b_{i\alpha s}$.

\begin{figure}
\begin{center}
\leavevmode
\epsfxsize = 80mm
\epsfbox{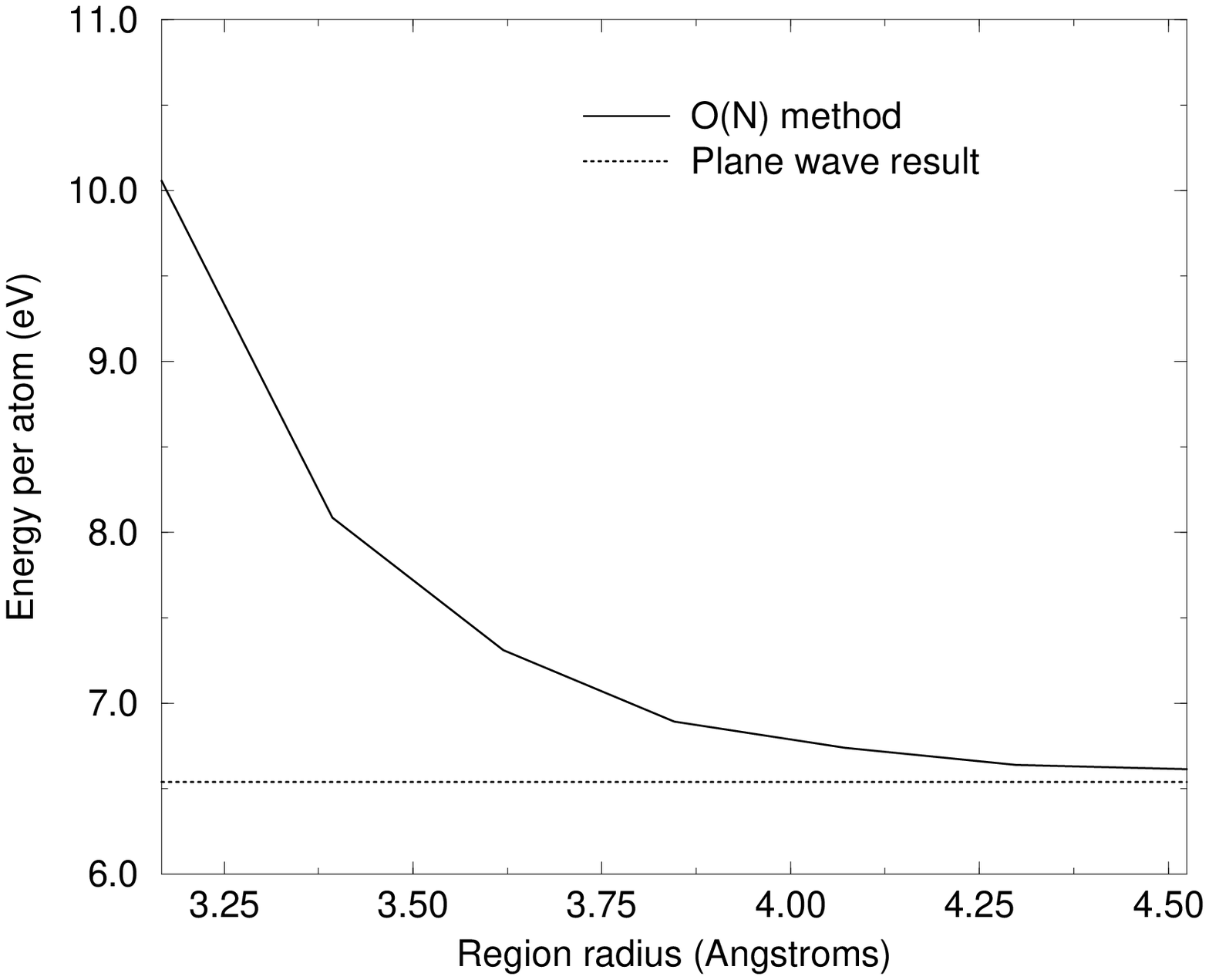}
\epsfxsize = 80mm
\epsfbox{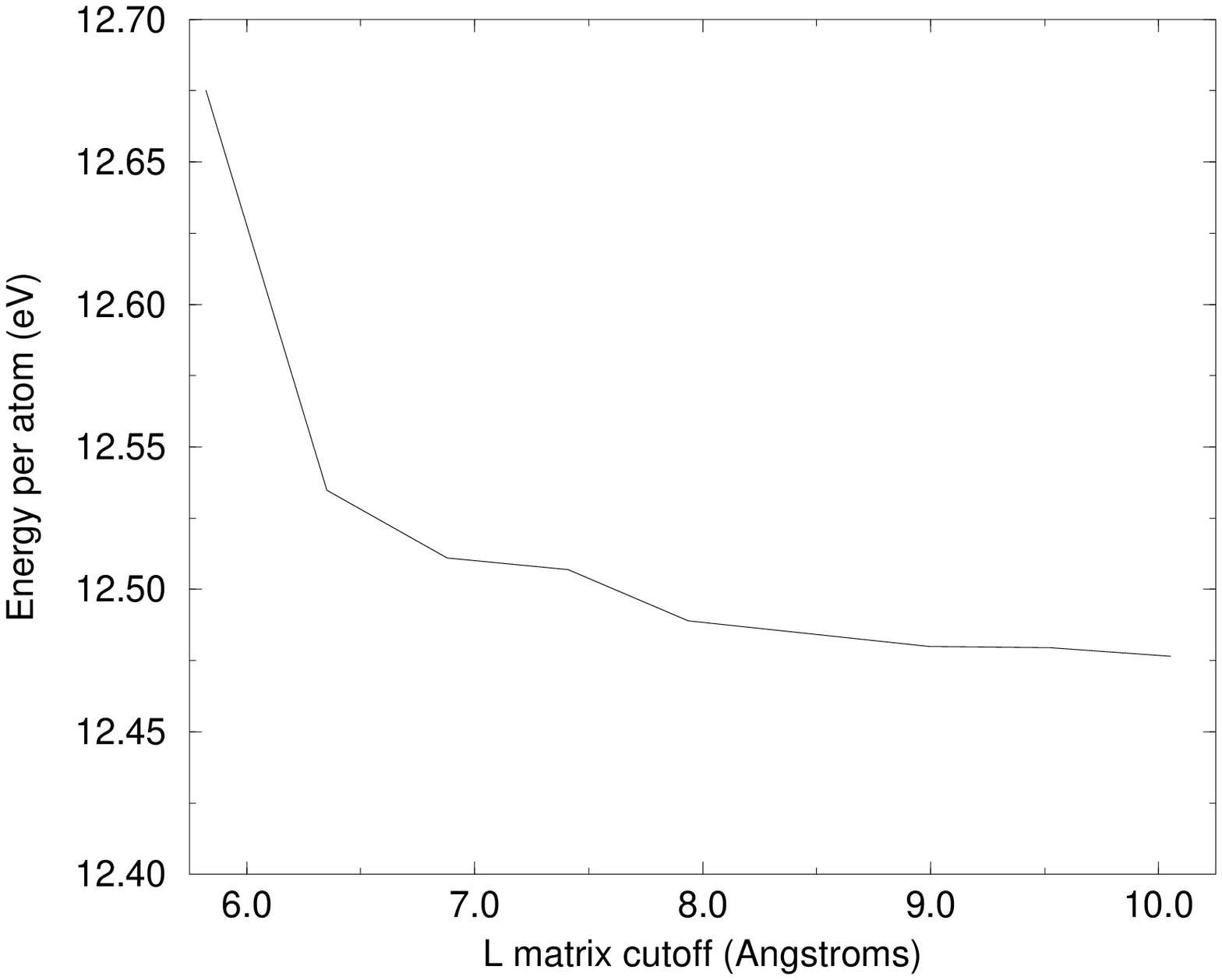}
\end{center}
\caption{Convergence to the true ground state energy of Si for 
\textsc{Conquest} as the cut-off applied $\rho$ is increased: 
(a) R$_{\rm reg}$ increasing with $R_L$ effectively infinite;
(b) $R_L$ increasing with R$_{\rm reg}$ held fixed.  See text for details.}
\label{fig:CQConv}
\end{figure}

Having described the minimisation of the total energy with respect to both the
$K$ matrix and the localised orbitals, it is now appropriate to consider the
practical performance of the method: are the cutoffs required to achieve
convergence to the ground state small enough to be practical ? Tests on a
model, local pseudopotential and standard non-local pseudopotentials have
already been reported\cite{Hern96,Gillan98}, and show that for reasonable
cutoffs, good convergence is obtained.  We reproduce some of these results in
Figure~\ref{fig:CQConv}.  Fig.~\ref{fig:CQConv}(a) shows the calculated total
energy as a function of $R_{\rm reg}$ for Si.  The results show that 
$E_{\rm tot}$ converges to the correct value extremely rapidly once 
$R_{\rm reg}$ is greater than ~4 \AA.  For this radius, each localisation 
region contains 17 neighbouring atoms, and the calculations are perfectly 
manageable.  Fig.~\ref{fig:CQConv}(b) shows the total energy for $R_{\rm reg}$
= 2.715 \AA, as a function of $R_L$.  Rather accurate convergence to
the $R_{\rm L} = \infty$ value is obtained for $R_{\rm L} \geq$ 8 \AA, which
again is acceptable.  No value is shown for exact diagonalisation because of 
technical difficulties in performing comparisons.

To perform integrations such as $S_{i\alpha j\beta} = \int {\rm d}{\bf r}
\phi_{i\alpha}({\bf r})\phi_{j\beta}({\bf r})$ numerical integration on
a grid is used.  This integration grid is generally of different spacing to
the blip grid (and normally about twice as fine).  Most matrix elements are
found by integration on this grid (with the exception of fast-varying
quantities which are calculated analytically), and the localised orbitals are
projected from the blip grid to the integration grid in a manner similar to a
fast Fourier transform (FFT), called a blip-to-grid transform.  
Once the charge density is calculated on the
grid (as $n({\bf r}) = \rho({\bf r},{\bf r})$), the Hartree potential and
energy are found using FFTs.

\textsc{Conquest} has been designed with parallel computers in mind; here we
summarize the strategy\cite{Goringe97}.
Each processor has three responsibilities.  First, a group of atoms: it
holds the blip coefficients, $b_{i\alpha s}$ and their derivatives of the 
energy, $\partial E_{\rm tot} /\partial b_{i\alpha s}$
and is responsible for performing the blip-to-grid transforms for these atoms.  
It also stores the rows of matrices corresponding to these atoms and 
performs the matrix multipications for these rows.  
Second, a domain of integration grid points: it has
responsibility for calculating contributions to matrix elements arising from 
sums over these points, and for holding the electron density and the Kohn-Sham
potential on these points.  Third, part of the spatial FFT for the Hartree
potential: it deals with a set of columns in the $x, y$ or $z$ directions.  
All processors switch between tasks in a concerted manner.  

To test the efficiency of the scheme, we have extensively tested its scaling
properties.  There are two completely different
kinds of scaling: parallel scaling
(i.e. scaling of CPU time for a {\it given} system with varying numbers of
processors); and inherent scaling (i.e. scaling of CPU time for a {\it fixed}
number of processors as the system size varies.  In the present implementation
of \textsc{Conquest}, both types of scaling are excellent\cite{Goringe97}.

The overall \textsc{Conquest} scheme can be summarised as follows: 
the ground state energy and
density matrix of the system are found by minimising the energy $E_{\rm tot}$
with respect to the elements of the auxiliary density matrix,
$L_{i\alpha,j\beta}$, and the localised orbitals, $\phi_{i\alpha}({\bf r})$,
subject to the spatial cutoffs $R_L$ and $R_{\rm reg}$. This yields an upper
bound to the true ground state, which improves as the cutoffs are increased.

\section{Strategies for reaching the ground state}
\label{Strat}

Having described the specific manner in which \textsc{Conquest} is
implemented, we now consider ways of reaching the ground state efficiently and
robustly.  At present, the minimisation is carried out in three separate 
stages: first, the ground state density matrix is found for a given set of 
localised orbitals; second, self-consistency is achieved between the charge 
density and the potential (which includes further density matrix minimisation
in response to the new potential); finally, the form of the localised orbitals 
is changed in accordance with the gradient of the energy.  The inner
loops (density matrix minimisation and self-consistency) are then repeated.
Schemes for efficiency and robustness for these three stages are now
discussed.

\subsection{Density matrix minimisation}
\label{sec:DMM}

\begin{figure}
\begin{center}
\leavevmode
\epsfxsize = 80mm
\epsfbox{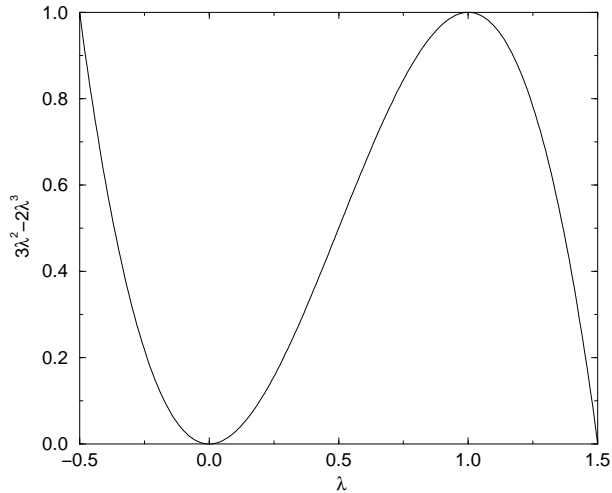}
\end{center}
\caption{The McWeeny purification function, $f(\lambda) = 3\lambda^2 -
2\lambda^3$.}
\label{fig:CubicL}
\end{figure}

As has already been emphasised, the density matrix of the true electronic 
ground state is idempotent.  This important property is hard to impose during a
minimisation; however, a number of schemes which drive the matrix 
towards idempotency have been suggested.  For simplicity, we will consider 
these schemes in the framework of orthogonal tight binding theory; the
extension to the non-orthogonal case and DFT is simple enough.
The first scheme was proposed by McWeeny\cite{McWeeny60}, who
noted that if a matrix $\rho$ is close to idempotency, then the matrix
$\tilde{\rho}$ given by:
\begin{equation}
\tilde{\rho} = 3\rho^2 - 2\rho^3
\label{eq:McW}
\end{equation}
will be more nearly idempotent.  It has the effect of driving 
the eigenvalues of $\rho$ towards zero and one (this can be seen by
considering the function $3\lambda^2 - 2\lambda^3$, which is shown in
Figure~\ref{fig:CubicL}).  If this transformation 
(often called the McWeeny transformation or purification transformation) is 
applied iteratively (writing $\rho_{n+1} = 3\rho^2_n - 2\rho^3_n$ for 
iteration $n+1$), then the sequence of matrices generated will converge on an 
idempotent matrix.   In fact, this transformation is {\it
quadratically} convergent (i.e. if the idempotency error in $\rho$ is $\delta
\rho$, then the idempotency error in $\tilde{\rho}$ is $\delta \rho^2$). 

Palser and
Manolopoulos\cite{Palser98} have recently suggested using this iterative
scheme in an ${\cal O}(N)$ manner.  They point out that if the initial 
density matrix is a linear function of the Hamiltonian, with
eigenvalues between zero and one, then the iteration will converge to 
the correct ground state density matrix (given by 
$\theta(\mu - H)$, where $\theta(x)$ is the Heaviside
step function ($\theta = 1$ for $x > 0$
and $\theta = 0$ for $x < 0$) and $\mu$ is the chemical potential for
electrons, or the Fermi energy), and that the energy, 
$E = 2{\rm Tr}[\rho_n H]$, will
decrease monotonically at each step.  This procedure has the
advantage of being fast (it only requires two matrix multiplies per
iteration) and efficient (it converges quadratically).  Unfortunately, when a
localisation criterion (also called a truncation) is applied to the density 
matrix to achieve linear scaling, the monotonic decrease of energy 
will fail at some point in the iterative search.  This can be 
taken as an indication that truncation errors are
dominating the calculation, and that the search should be
stopped\cite{Palser98}; indeed, if it is not stopped, there is no guarantee
that it will continue to converge towards an idempotent matrix.
This is a heuristic criterion for stopping the iteration, and has the
drawback that the method will not be variational, so that analytic 
forces will not be in agreement with the numerical gradient of the energy.

The Li, Nunes and Vanderbilt (LNV) scheme for achieving the ground state density
matrix\cite{(Li etal 1993),NV} also uses the McWeeny transformation, though
in a rather different manner.  Here, the energy is written as $E = 2{\rm
Tr}[\tilde{\rho} H]$, with $\tilde{\rho}$ given by equation~\ref{eq:McW}. Then
the energy is minimised with respect to the elements of $\rho$, typically
using a scheme such as conjugate gradients\cite{NR} to generate a sequence of
search directions.  The localisation of the density matrix is achieved by
applying a spatial cutoff to the elements of $\rho$.  This scheme has at 
least two advantages: first, each line
minimisation can be performed analytically, as the energy is cubic in $\rho$;
second, it is variational, so that the energy found is always an upper bound
to the ground state, and forces obey the Hellmann-Feynman theorem and are in
exact agreement with the numerical derivative of the energy.

However, there are drawbacks to the LNV technique.  
First, it is unclear what the best 
initial value should be for the density matrix; typically, it is taken to be
$\frac{1}{2}${\bf I}, or $\frac{1}{2}${\bf S$^{-1}$} in a non-orthogonal basis
set.  Second, as the McWeeny transformation is a cubic, it is unbounded from
below, and a poor starting choice for the minimisation can lead to runaway
solutions; a sign of this is typically that the cubic has complex extrema.
Third, the scheme can be poorly convergent, and is not guaranteed the
quadratic convergence of the McWeeny method.

We have recently proposed a hybrid between the McWeeny and LNV schemes which
builds on the complementarity between these two methods\cite{Bowler99}.  
It is based
on the observation that the sequence of matrices generated during a McWeeny
iterative search converges quadratically towards idempotency, and that the LNV
search direction maintains idempotency in the density matrix 
to first order.  Thus the McWeeny scheme is used as an initialisation 
to find an idempotent density matrix (but one which is not the ground 
state matrix because of truncation errors); this matrix
is then used as the input to the LNV scheme, which maintains the 
idempotency while searching efficiently for the ground state density matrix.  
The combination of the two methods is both variational and
robust - two highly desirable attributes for the inner loop of an ${\cal
O}(N)$ DFT method.

\begin{figure}
\begin{center}
\leavevmode
\epsfxsize = 80mm
\epsfbox{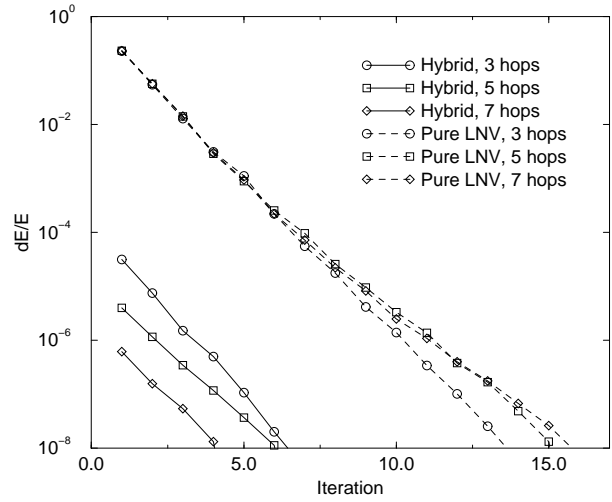}
\end{center}
\caption{The difference between the cohesive energy at a given iteration and
the final cohesive energy for the LNV stage of the hybrid scheme (solid lines)
and the pure LNV scheme (dashed lines) for diamond structure carbon.  
Results are shown for different cut-off radii: 3 hops (circles), 
5 hops (squares) and 7 hops (diamonds), where a hop is a Hamiltonian nearest
neighbour distance.}
\label{fig:McWConv}
\end{figure}

As an example of the improved speed of convergence given by the hybrid scheme,
Figure~\ref{fig:McWConv} shows the convergence to the ground state energy in
diamond carbon for the LNV stage of the hybrid scheme and pure LNV
(initialised from $\rho = \frac{1}{2}{\bf I}$).  These results show that the
McWeeny stage of the hybrid scheme gets closer to the ground state as the
radius is increased, as expected, and that it acts as an excellent initial
density matrix for the LNV scheme.  The method has also been tested on a
vacancy in diamond C, the Si(001) surface and liquid Si, as reported
elsewhere\cite{Bowler99}.

\subsection{Non-orthogonality}

The previous section focuses on the traditional tight binding scheme
where the localised orbitals are taken to be orthogonal.  However, in our
${\cal O}(N)$ method, the orbitals are non-orthogonal, and this
introduces a significant degree of complication.  In strict mathematical
terms, the metric for the space spanned by the localised orbitals must be
chosen with care; this means defining a scalar product in a specific way, as
explained in detail in Appendix \ref{App:NO}, along with other implications
of using non-orthogonal orbitals.  We will consider a few simple implications 
of this theory here.

For any given set of non-orthogonal orbitals, \{$\bar{\phi}_i$\}, 
an orthogonal set can be defined by using the overlap matrix, $S$:
\begin{equation}
\phi_i = \sum_j (S^{-1/2})_{ij} \bar{\phi}_j,
\end{equation}
where we use an over-bar to indicate the quantities in the non-orthogonal
case.  If the metric is chosen suitably, then similar transformations between 
the Hamiltonian and density matrices in the two spaces can be defined:
\begin{eqnarray}
\bar{H} & = & {\bar{S}}^{1/2} H {\bar{S}}^{1/2} \nonumber \\
\bar{\rho} & = & {\bar{S}}^{- 1/2} \rho {\bar{S}}^{- 1/2} \; .
\label{eq:OrthoNO}
\end{eqnarray}
There are various points which can be drawn from the above equations.  First,
the McWeeny transformation (and the other quantities associated with a
minimisation such as the gradient of the energy) will change in the new basis
set; in fact the McWeeny transformation becomes $\tilde{\bar{\rho}} = 
3 \bar{\rho} S \bar{\rho} - 2 \bar{\rho} S \bar{\rho} S \bar{\rho}$.  
Second, there are different types of matrix in the
non-orthogonal case, one of which transforms with ${\bar{S}}^{1/2}$ and 
the other with ${\bar{S}}^{- 1/2}$ (in fact there are two types of orbital and 
hence four types of matrix); great care must be taken to combine these
matrices in the correct fashion, as pointed out by White {\it et
al.}\cite{White97} for the case of the gradient of the energy with the
non-orthogonal density matrix.  Third, there may be a considerable advantage
in choosing the metric so that the transformations shown in equation
\ref{eq:OrthoNO} apply, as this will enable direct comparison with the
orthogonal case.  The interested reader is referred to
Appendix~\ref{App:NO} and references therein for more details.

\subsection{Charge mixing and self-consistency}

The question of achieving self-consistency between the charge density and the
potential has been examined in great detail over many years, and much is known
about efficient implementation\cite{Kresse96}.  Within \textsc{Conquest}, the 
direct inversion of the iterative subspace (DIIS) method of 
Pulay\cite{Pulay} is used.  At each iteration, a residual can be defined as:
\begin{equation}
R[\rho^{\rm in}] = \rho[\rho^{\rm in}] - \rho^{\rm in},
\label{eq:Resid}
\end{equation}
where $\rho[\rho^{\rm in}]$ is the output charge density: that is, the potential
arising from $\rho_{\rm in}$ is found (consisting of Hartree and
exchange-correlation parts), the Schr\"odinger equation is solved, and the
output charge density generated from the resultant wavefunctions. 
Clearly, the aim is to reduce $R[\rho^{\rm in}]$ to as close to zero as
possible in the least number of iterations.  
The simplest possible technique is to use the output charge density from one
cycle ($\rho^{\rm out}_{n} = \rho[\rho^{\rm in}_n]$) as the input for the
next cycle: $\rho^{\rm in}_{n+1} = \rho^{\rm out}_{n}$; however, 
this is potentially rather slow, and prone to the
phenomenon known as `charge sloshing', where long wavelength variations of the
charge in the unit cell dominate the self-consistency procedure.  There are in
fact cases where this simple method fails to work at all, and self-consistency
is never reached.  This is clearly unacceptable.

Better than this is to perform a linear mix of the two previous charge
densities, so that:
\begin{equation}
\rho^{\rm in}_{n+1} = (1-\lambda)\rho^{\rm out}_{n-1} + 
\lambda\rho_n^{\rm out}.
\end{equation}
The value of lambda can be found easily.  If the residuals (defined above in
equation~\ref{eq:Resid}) are treated as vectors (considering the value at each 
spatial position as
an entry in the vector), then scalar products can be formed between residuals,
and the norm of a residual can be found as $\left( \langle 
R[\rho_{n+1}^{\rm in}] \mid R[\rho_{n+1}^{\rm in}] \rangle \right)^{1/2}$.  The
optimum value of $\lambda$ is found by minimising this norm with respect to
$\lambda$\cite{Johnson}.  This gives:
\begin{equation}
\lambda = { \langle R[\rho^{\rm in}_n] \mid R[\rho^{\rm in}_n] - R[\rho^{\rm
in}_{n-1}]\rangle \over \langle R[\rho^{\rm in}_n] - R[\rho^{\rm in}_{n-1}] 
\mid R[\rho^{\rm in}_n] - R[\rho^{\rm in}_{n-1}] \rangle}.
\end{equation}
This procedure can be generalised to more than two previous densities, which
can give significant benefits, as described elsewhere\cite{Kresse96,Pulay}.
It is often important to mix in a small amount of the input charge densities,
as well as performing the mixing given above.

As well as combining charge densities in the optimum manner, it is important 
to suppress the phenomenon of `charge sloshing'; an ideal way to do this is to
use Kerker preconditioning\cite{Kerker}.  Here a scaling is applied
in reciprocal space to the residual:
\begin{equation}
R[\rho^{\rm in}_j] = R[\rho^{\rm in}_j]\times {q^2 \over q^2 + q^2_0},
\end{equation}
where $q$ is a reciprocal space vector, and $q_0$ is chosen suitably (a value
close to $2\pi / a_0$, where $a_0$ is a lattice vector, is appropriate).
This scaling is an approximation to the inverse dielectric function, and 
enables fast and robust iteration to a self-consistent charge density 
and potential.

\subsection{Pre-conditioning localised orbital variation}

Now that we have described the robust and efficient search for the ground 
state density matrix (for a given set of localised orbitals) and the fast
iteration to a self-consistent charge density and potential, we must consider
the variation of the localised orbitals.  

\begin{figure}
\begin{center}
\leavevmode
\epsfxsize = 80mm
\epsfbox{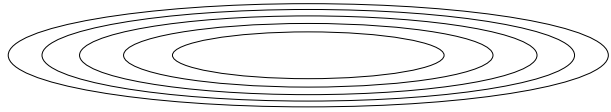}
\end{center}
\caption{A function with elongated surfaces of constant $f$}
\label{fig:curve}
\end{figure}

As is the case for minimisation problems in many areas of science, 
\textsc{Conquest} suffers from
ill conditioning in the search for the ground state when varying the localised
orbitals.  Ill conditioning occurs if the function being minimised has a wide
range of curvatures.  For a general function $f(x_1, x_2,...,x_N)$, dependent
upon variables $\{x_i\}$, the curvature matrix (or Hessian) can be defined as
$C_{ij} = \partial^2 f/\partial x_i \partial x_j$.  If the eigenvalues
$\lambda_n$ of $C_{ij}$ span a wide range, then the surfaces of constant
$f$ are elongated (illustrated in Figure~\ref{fig:curve}), and 
conventional techniques such as conjugate
gradients\cite{NR} will become very inefficient.  It is known that the
number of iterations required by conjugate gradients is proportional to
$(\lambda_{\rm max}/\lambda_{\rm min})^{1/2}$, where $\lambda_{\rm max}$ and
$\lambda_{\rm min}$ are the maximum and minimum eigenvalues of
$C_{ij}$\cite{Annett95}.

While ill conditioning is a widespread problem, the solution depends on the
specific situation.  Conventional first principles calculations which use
plane waves as a basis set have been recognised for many years to suffer from
ill conditioning\cite{Payne92,Gillan89}, and it turns out that ill
conditioning found in ${\cal O}(N)$ calculations is closely related to this;
we will review the plane wave ill conditioning before describing the ${\cal
O}(N)$ ill conditioning.

The plane wave energy functional (eq.~\ref{eq:EtotPW}) has large curvatures 
associated with high wavevector {\bf G}, because of the form of the 
kinetic energy:
\begin{equation}
E_{\rm kin} = 2\sum_i f_i \sum_{\bf G}{\hbar^2 G^2 \over 2m}\mid c_{i{\bf G}}
\mid^2,
\end{equation}
so that the energy is proportional to $G^2$.  This first type of ill
conditioning is easily cured (essentially by scaling $c_{i{\bf G}}$ by a
factor of $(1+G^2/G^2_0)^{-1/2}$), and is referred to as `length scale ill
conditioning', as it comes from the variation of curvature with length scale.

Another type of ill conditioning seen in conventional techniques is associated
with the invariance of $E_{\rm tot}$ under a unitary transformation of the
orbitals.  If the occupation numbers $f_i$ are all either zero or one, then
$E_{\rm tot}$ is exactly invariant under transformations such as:
\begin{equation}
\psi_i^\prime = \sum_j U_{ij} \psi_j,
\end{equation}
where $U_{ij}$ is a unitary matrix.  If the occupancies deviate slightly from
zero or one, however, the exactness of the invariance is broken, and the
energy changes slightly.  Some of the eigenvalues of the Hessian will go from
exactly zero (under the exact transformation) to very small, leading to poor
conditioning.  We shall refer to this as `superposition ill conditioning'.  In
conventional techniques this is cured by performing a rotation of the wave 
functions so that the Hamiltonian becomes diagonal in the subspace spanned by
the occupied states.

A final type of ill conditioning found in conventional methods arises with
variable occupation numbers, and is associated with eigenvalues whose 
energies are well above the Fermi energy, which will have very small
occupation numbers.  Variations of the $\psi_i$ associated with these small
occupation numbers will have little effect on the value of $E_{\rm tot}$, and
lead to small values of the curvature.  Since the variations of these
eigenvalues are almost redundant in the minimisation, we refer to this as
`redundancy ill conditioning'.

\begin{figure}
\begin{center}
\leavevmode
\epsfxsize = 80mm
\epsfbox{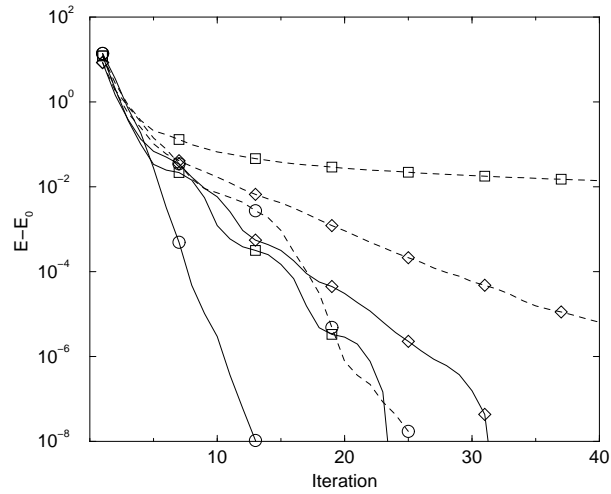}
\end{center}
\caption{Convergence to the ground state of the Si crystal calculated with 
(solid lines) and without (dashed lines) preconditioning.   Results are for
region radii of 2.72~\AA\ (circles), 3.40~\AA\ (squares) and 3.80~\AA\
(diamonds).}
\label{fig:LSIC}
\end{figure}

All three of these forms can cause difficulties within ${\cal O}(N)$
techniques, though typically the specific solution will vary.  For instance,
it is clear that variations of the localised orbitals, $\{\phi_{i\alpha}\}$, 
will have different length scales, and will suffer from length scale ill
conditioning.  This is easily cured in the same way as for plane waves, as has
been recently demonstated\cite{Bowler98}, though the methodology is somewhat
different.  As a demonstration of the efficacy of this preconditioning, 
Figure~\ref{fig:LSIC} shows the convergence of \textsc{Conquest} with
and without length scale preconditioning for three different region radii for
the localised orbitals.  Clearly this problem becomes significantly worse 
for larger regions.

Superposition ill conditioning is associated with the linear mixing of
localised orbitals.  It is easily shown that linear mixing of the functions 
on the same atom leaves $E_{\rm tot}$ unchanged, and so will not cause 
ill conditioning.
Variations of the localised orbital $\phi_{i\alpha}$ such as:
\begin{equation}
\phi_{i\alpha}^\prime =  \phi_{i\alpha} + 
\sum_{j\beta, j\ne i} c_{j\beta} \phi_{j\beta}
\end{equation}
are rather more interesting.  Strictly, these are not possible, as the 
localised orbitals are constrained to be zero outside their localisation 
regions.  However,
once the region radii become large, there will be variations which almost
fulfil this criterion.  It is the small eigenvalues of the Hessian of $E_{\rm
tot}$ associated with this almost perfect mixing which will cause superposition 
ill conditioning.  It is perfectly possible to cure this, however, as the form
of the variations can be written down.  We have developed a method to
precondition these variations, and are testing it.  It will be described in a
future publication.

Finally, we come to redundancy ill conditioning.  Just as in conventional
calculations this occurs when the occupation numbers $f_i$ are very small,
this may occur in ${\cal O}(N)$ techniques when the number of localised
orbitals $\phi_{i\alpha}$ is more than half the number of electrons.  It is
desirable, if not essential, to be able to work with an extended number of
orbitals; for instance, in group IV elements, the natural basis will consist
of four orbitals, roughly corresponding to one $s$ and three $p$ orbitals.
(It is worth noting that Kim, Mauri and Galli\cite{Kim96} have found that it
is {\it essential} to have more orbitals than filled bands to avoid local
minima in the energy functional in a related ${\cal O}(N)$ scheme.)  As before,
we believe that this form of ill conditioning can be removed by suitable
preconditioning, but detailed techniques have yet to be formulated.

\section{Future directions}
\label{Fut}

Having summarised the techniques involved in \textsc{Conquest}, it is 
appropriate to look to the near future, and consider the directions in which
the project is going.  

\subsection{Forces}
The issue of forces is a key one for any electronic structure technique; if
relaxation of ions or molecular dynamics are to be performed then the
analytical forces must agree with the gradient of the energy.
\textsc{Conquest} has been specifically constructed so that, provided small
Pulay-type corrections\cite{Pulay69} are included, the forces are guaranteed
to be consistent with the energy.  Pulay corrections are required because the
B-spline basis functions move with the atoms, and are easily found, as will be
described elsewhere.  This means that the relaxation of the system to
mechanical equilibrium and the generation of time-dependent ionic 
trajectories will be feasible in ${\cal O} (N)$ DFT calculations.

\subsection{Efficient choices for representation of localised orbitals}

The present choice of basis for representing the localised orbitals has been
described in Section~\ref{sec:CQ}.
However, there are two good reasons for changing this basis somewhat.  First,
there is the problem of ripples in the energy caused by the numerical
integration; this is due to a lack of translational invariance with respect
to the integration grid.  If the rapidly varying parts of the 
localised orbitals (i.e. the
core regions, which do not alter greatly during a calculation) could be
represented in a more efficient form, then the integrals could be performed
analytically, significantly reducing ripples.  Second, there would be great
value in being able to perform {\it ab initio} tight binding calculations with
the code (or even to have certain parts of a supercell treated with full DFT,
while others were treated with {\it ab initio} tight binding).  For these
reasons, we are planning to move to a mixed basis (seen recently in
conventional pseudopotential calculations) which combines pseudo-atomic
orbitals (possibly of the form of Sankey and Niklewski\cite{SN}) with a
coarser blip grid.  This will suppress the ripples with respect to the
integration grid, and give the flexibility to model different parts of the
system with appropriate accuracy.

\subsection{Finding the density matrix for metals}

The question of modelling metals is much harder than insulators or
semiconductors for the ${\cal O}(N)$ methods
described above, simply because the density matrix is more delocalised in
metals, meaning that the cutoff applied to $L$ (and hence to $K$) 
must be much larger to obtain accurate results.
If the metal is close packed, as is frequently the case, this entails a rapid
increase in the number of elements in the density matrix, and a slowing down
of variational techniques; this is discussed elsewhere\cite{Bowler97}.

One approach is to reduce the range of the density matrix in metals 
by introducing an artificial
electronic temperature, which broadens the Fermi occupation function, and
localises the density matrix.  The drawback is a potentially large electronic
entropy contribution to the energy; however, there is a scheme for
extrapolating the results back to zero electronic temperature\cite{Gillan89}.
It has been shown\cite{Bowler97} that even with such a scheme the
variational density matrix method described above is inefficient; the hybrid
method described in Section \ref{sec:DMM} improves the
efficiency\cite{Bowler99}.
More efficient are recursion-based methods, such as the Fermi Operator
Expansion\cite{Goe94} or the Bond Order Potential\cite{Pett89,Aoki93}.
However, these have the disadvantage of not being strictly variational.  

A further possibility which has emerged recently is to use a series of
nested Hilbert spaces to find the exact zero temperature
density matrix\cite{Baer98}.  This method has the advantage that no
approximation is being made to remove the electronic entropy contributions,
and allows high precision calculations on metals.  It is, however, still in
development, and practical demonstrations have yet to be published.

\section{Conclusions}
\label{Conc}
The recent developments which have been outlined above show that the future of
${\cal O}(N)$ {\it ab initio} techniques is extremely 
bright.  We have shown that the
localisation of the density matrix gives the framework within which these
methods can be constructed, and have given details of the implementation
of one such code, \textsc{Conquest}.  The examples presented show that this
method is practical, and that the spatial cutoffs required for accuracy 
are small enough to make the calculations perfectly feasible.  
The search for the ground state has been addressed, and methods for making
this search more robust and efficient have been discussed.  The remaining tasks
for the \textsc{Conquest} code have been described, and the way forward for all
of them is clear.  The most important conclusion to draw
from this body of work is that ${\cal O}(N)$ DFT methods actually work. 
Indeed, these methods are being demonstrated in practical
calculations.  Our group is working towards practical application of the
\textsc{Conquest} code to large-scale problems.

\section*{Acknowledgements}
We are happy to acknowledge useful discussions with David Manolopoulos, Peter
Haynes, Chris Goringe and Ed Hern\'andez.  
\textsc{Conquest} has been developed within the framework of the UK 
Car-Parrinello consortium, which is supported by the EPSRC grant
GR/M01753.  The work of MJG is financially supported by CCLRC and GEC.

\appendix

\section{The pseudopotential method}
\label{sec:Pseudo}

As \textsc{Conquest} uses the standard pseudopotential method, it is
important to recall the salient facts within this formalism; there are many
excellent reviews elsewhere\cite{Payne92} which give more detail.  

In the pseudopotential method, only the valence electrons are considered, and
their interaction with the ionic cores is replaced by a {\it pseudopotential}, 
$v(r)$.  This means that in fact the solutions of the Schr\"odinger equation
are pseudo-wavefunctions, and the charge density, $n({\bf r})$, is the
pseudo-density of the valence electrons.  The energy arising from the
interaction between the cores and the valence electrons is given by:
\begin{equation}
E_{ei} = \int {\rm d}{\bf r}V({\bf r})\rho({\bf r}),
\end{equation}
where $V({\bf r})$ is found as the sum over the ionic pseudopotentials:
\begin{equation}
V({\bf r}) = \sum_i v(\mid {\bf r}-{\bf R}_i\mid),
\end{equation}
with ${\bf R}_i$ the core positions.  In general, the pseudopotential is
non-local, that is $v({\bf r},{\bf r}^\prime)$.

In conventional pseudopotential {\it ab initio} techniques, the wavefunctions
are often expanded in terms of plane waves:
\begin{equation}
\psi_i = \sum_{\bf G} c_{i{\bf G}} {\rm exp}(i{\bf G}\cdot{\bf r}),
\end{equation}
where {\bf G} is a reciprocal lattice vector.
The total energy is then minimised with respect to the coefficients $c_{i{\bf
G}}$.  Frequently, particularly in metals, variable occupation numbers are 
allowed, so that the wave function $\psi_i$ has an occupation $f_i$.  
This gives a total energy function: 
\begin{equation}
E_{\rm tot} = E_{\rm tot}(\{c_{i{\bf G}}\},\{f_i\}).
\label{eq:EtotPW}
\end{equation}

\section{Metrics and minimisation in non-orthogonal basis sets}
\label{App:NO}

When working with a non-orthogonal basis, care must be taken with notation.
It is common to use raised and lowered indices to distinguish between vectors
and matrices which transform differently; this has been introduced to
electronic structure calculations by Ballentine and Kol\'a\v{r}\cite{BK86}, 
who also describe the general formalism.

The eigenfunctions of the Hamiltonian are expanded in a set of non-orthogonal, 
localised, atom-centred orbitals $\{ \phi_\alpha({\bf r}) \}$, 
where $\alpha$ runs over all orbitals on all atoms.  These orbitals 
span a Hilbert space ${\cal V}$, and define an overlap matrix which 
is given by $S_{\alpha \beta} = \langle \phi_\alpha \mid \phi_\beta \rangle$.
The inverse overlap matrix, $\left( S^{-1} \right)^{\alpha \beta}$, is 
defined by the relation:
\begin{eqnarray}
\sum_\beta \left( S^{-1} \right)^{\alpha \beta} S_{\beta \gamma} &=&
\delta_\gamma^\alpha\nonumber\\
 &=& 1 \hskip 10mm {\rm if} \alpha = \gamma \nonumber\\
 &=& 0 \hskip 10mm {\rm if} \alpha \neq \gamma.
\end{eqnarray}

A {\it dual space}, ${\cal V^*}$, exists which is spanned by the 
orbitals $\mid \phi^\alpha \rangle = \sum_\beta \left( S^{-1} \right)^{\alpha
\beta} \mid \phi_\beta \rangle$.  These two sets of vectors are {\it
bi-orthogonal}, that is $\langle \phi^\alpha \mid \phi_\beta
\rangle = \delta^\alpha_\beta$.  It is important to note that contraction can 
only be carried out over indices which are opposed, while addition 
can only be carried out between tensors for which indices agree.  The
vectors in the original space are called {\it covariant}, while the vectors in
the dual space are {\it contravariant}.

A convenient choice for the metric of ${\cal V}$ is $S^{-1}$, so that 
$\sum_\beta \langle \phi_\alpha \mid \left(S^{-1}\right)^{\gamma \beta} 
\mid \phi_\beta \rangle = \delta_\alpha^\gamma$.  An equivalent choice of 
metric for ${\cal V^*}$ is $S_{\alpha \beta}$.  These operate to change 
a vector in one space to
the vector in another; thus a proper scalar product within a space can be
formed by incorporation of the metric.  A covariant operator can be represented 
as an {\it outer product}:
\begin{equation}
\hat{A} = \sum_{\alpha,\beta} \mid \phi_\alpha \rangle A^{\alpha \beta} \langle
\phi_\beta \mid.
\end{equation}
Then the scalar product of two covariant operators is written:
\begin{eqnarray}
({\bf A},{\bf B}) &=& \sum_{\alpha,\beta,\gamma,\delta}\langle \phi_\delta 
\mid \phi_\beta \rangle A^{\beta \alpha} \langle
\phi_\alpha \mid \phi_\gamma \rangle B^{\gamma \delta}, \nonumber \\
 &=& \sum_{\alpha,\beta,\gamma,\delta} S_{\delta \beta} A^{\beta \alpha} 
S_{\alpha \gamma} B^{\gamma \delta}\nonumber \\
 &=& Tr[A^{\dag}SBS].
\end{eqnarray}

It is important to note that the product of a covariant and a contravariant 
pair (such as $H$ and $\rho$) is
invariant with basis set.  Traditionally, the Hamiltonian is taken as
covariant and the density matrix as contravariant.

\subsection{Variation of {\it L}}

The innermost part of \textsc{Conquest} consists of the minimisation 
of the energy with
respect to the elements of the matrix $L^{i\alpha j\beta}$ with fixed support
functions.  This is achieved by performing line minimisations along directions
supplied by the conjugate gradients algorithm (with, on occasion, a correction
for maintaining the electron number constant).  As pointed out recently by 
White {\it et al.}\cite{White97}, the gradient of the energy with respect to 
$L^{i\alpha j\beta}$, $\nabla \Omega$, (which is used as the search direction 
in the minimisation) is actually {\it covariant}, while $L^{i\alpha j\beta}$ is 
{\it contravariant}; this means that the gradient must be transformed to a
contravariant matrix before being combined with the density matrix:
$(S^{-1})\nabla \Omega(S^{-1})$.  
But there is more to the problem of minimisation than just 
this; conjugate gradients assumes the following relations:
\begin{eqnarray}
{\bf g}_{i+1} &=& -\nabla f({\bf P}_{i+1}), \nonumber\\
{\bf h}_{i+1} &=& {\bf g}_{i+1} + \gamma_i{\bf h}_i, \nonumber\\
\gamma_i &=& {{\bf g}_{i+1} \cdot {\bf g}_{i+1} \over {\bf g}_i \cdot {\bf
g}_i},
\end{eqnarray}
where ${\bf g}_i$ is the gradient of the function at step $i$ and ${\bf h}_i$
is the search direction (which is conjugate to the previous search
directions) at step $i$.  Clearly in the formation of $\gamma_i$, care must be
taken to ensure that the product is tensorially correct, otherwise the choice
of the new search direction will be wrong, so the correct formula for 
$\gamma_i$ is:
\begin{equation}
\gamma_i = {{\bf g}_{i+1} \cdot (S^{-1}) {\bf g}_{i+1} (S^{-1}) \over 
{\bf g}_i \cdot (S^{-1}) {\bf g}_i (S^{-1})},
\end{equation}
(This is discussed at more length in Section 2.7 of Ref.\cite{NR}, where 
the {\it bi-conjugate} gradient method is described.  This degree of 
complexity is not needed here.) 
Similar care both with the correct nature of gradients and with the search
directions must be taken when varying the localised orbitals.

\end{multicols}

\widetext

\end{document}